\documentclass[sigconf]{acmart}

\AtBeginDocument{%
  }

\usepackage{algorithm}
\usepackage{algpseudocode}
\usepackage{amsmath}
\usepackage{float}
\usepackage{booktabs}
\usepackage{graphicx}
\usepackage{placeins}

\setcopyright{acmlicensed}
\copyrightyear{2025}
\acmYear{2025}
\acmConference[ICAIF '25]{6th ACM International Conference on AI in Finance}{November 15--18,
  2025}{Singapore}
\begin{document}
\raggedbottom

\title{Regret-Driven Portfolios: LLM-Guided Smart Clustering for
Optimal Allocation}

\author{Muhammad Aarash}
\affiliation{%
  \institution{Lahore University of Management Sciences}
  \city{Lahore}
  \country{Pakistan}}
\email{25100330@lums.edu.pk}

\author{Dr. Hassan Jaleel}
\affiliation{%
  \institution{Lahore University of Management Sciences}
  \city{Lahore}
  \country{Pakistan}}
\email{hassan.jaleel@lums.edu.pk}

\begin{abstract}
We attempt to mitigate the persistent tradeoff between risk and return in medium- to long-term portfolio management. This paper proposes a novel \textbf{LLM-guided no-regret portfolio allocation framework} that integrates online learning dynamics, market sentiment indicators, and large language model (LLM)-based hedging to construct high-Sharpe ratio portfolios tailored for risk-averse investors and institutional fund managers. Our approach builds on a follow-the-leader approach, enriched with sentiment-based trade filtering and LLM-driven downside protection. Empirical results demonstrate that our method outperforms a SPY buy-and-hold baseline by \textbf{69\%} in annualized returns and \textbf{119\%} in Sharpe ratio.
\end{abstract}




\maketitle

\section{Introduction}
\label{sec:intro}
From 2014 to 2024, only about 6.6\%\cite{GehringerPauli2025} of hedge funds outperformed the S\&P 500 on a nominal return basis, with many professional fund managers delivering inferior returns despite higher fees. A notable illustration of this trend is Warren Buffett’s 2007 wager\cite{Buffett2017Letter} that a low-cost S\&P 500 index fund would beat a portfolio of hedge funds over ten years – a bet he handily won in 2017 (85.4\% total return vs. the hedge funds’ 22\%\cite{PerryAEI}). This shift in fortunes has intensified the spotlight on passive investing and raised doubts about traditional active stock-picking strategies. 

However, nominal outperformance is only part of the picture. Institutional and individual investors alike are risk-conscious, often prioritizing risk-adjusted returns over raw returns. For example, pension and retirement fund managers seek strategies that offer downside protection and low correlation to market swings, since these portfolios represent individuals’ life savings. Such an emphasis on hedging typically leads to lower active share \footnote{the percentage of holdings in a fund’s portfolio that differ from its benchmark index} and reduced upside potential. Managers also tend to rebalance infrequently to minimize transaction costs and avoid alarming investors. In general, higher returns come with higher volatility – a fundamental risk–return trade-off. This reality poses a challenge: How can we achieve superior returns without incurring undue risk or frequent trading? 

In light of these challenges, we propose an \textbf{LLM-Guided “no-regret” portfolio allocation framework}\footnote{Source code available at: \href{https://anonymous.4open.science/r/no-regret-paper-62CA/README.md}{anonymous.4open.science/r/no-regret-paper-62CA}} that adaptively balances return and risk. Our approach is formulated as a sequential decision-making problem using \textit{no-regret online learning principles}. The strategy learns to minimize regret relative to the best static portfolio in hindsight, ensuring that over time its performance approaches that of the optimal benchmark portfolio. To keep the solution tractable, we constrain the search space to incremental adjustments around a diversified baseline portfolio, this preserves a wide range of possible allocations while limiting extreme shifts. 

\begin{figure}
  \centering
  \includegraphics[width=0.45\textwidth]{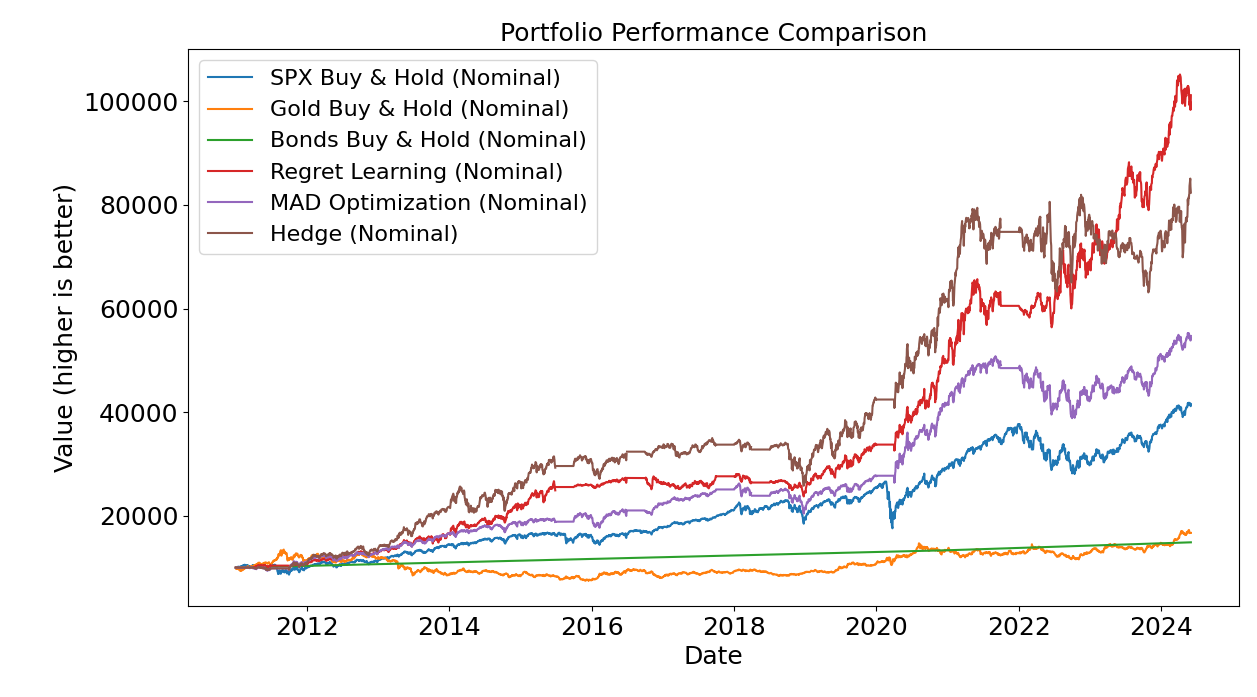}
  \caption{\small{The No Regret Learning strategy (red) significantly outperforms buy-and-hold benchmarks (SPX, gold, bonds) within the sector-based ETFs asset universe, demonstrating lower volatility and resilience during major market downturns (e.g., 2020 COVID-19 crash, 2022 inflation shock) than benchmarks like SPY, as well as online approaches like MAD (online) \cite{Konno1991} and Hedge (multiplicative weights) \cite{Helmbold1998}.}}
  \label{nrlperformance}
  \vspace{-0.1in}
\end{figure}
% increase font (labels, legends, caption) + make dotted lines gray and thinner

\begin{figure}
  \centering
  \includegraphics[width=0.4\textwidth]{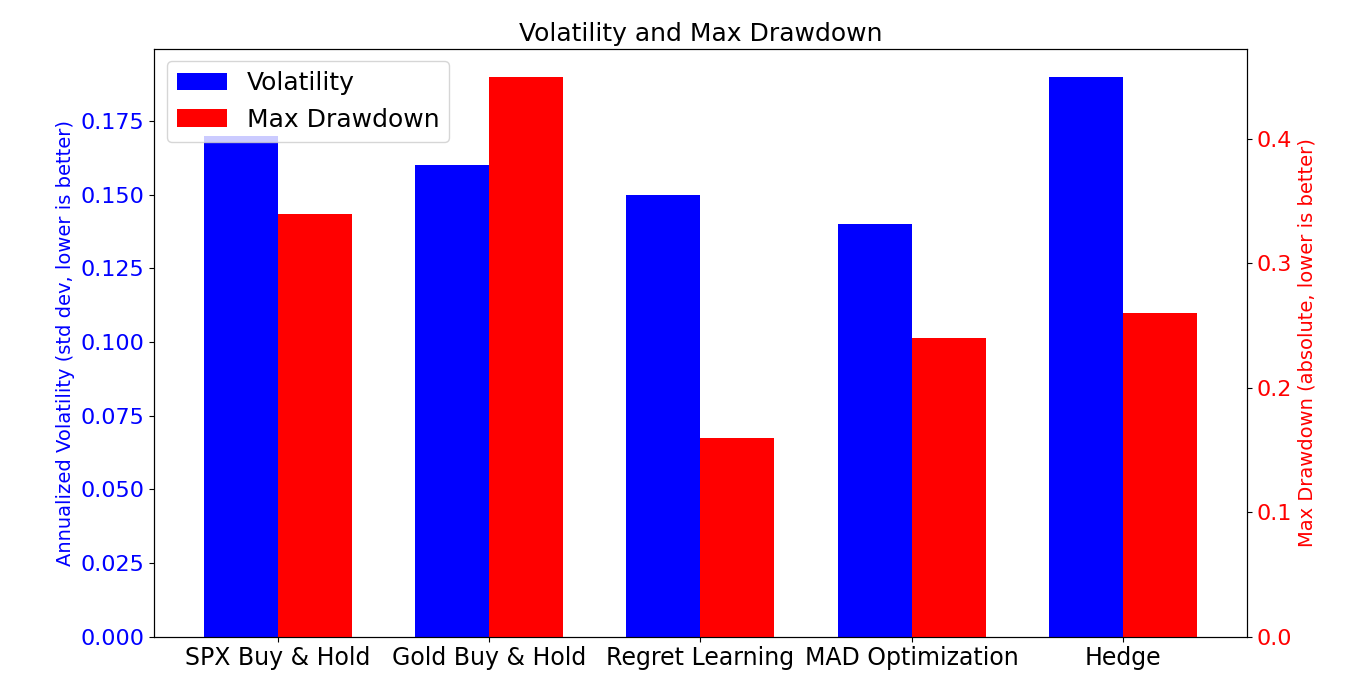}
  \caption{\small{The volatility and drawdown statistics for the NRL tests in Figure \ref{nrlperformance} (Jan 2011 - June 2024). Here we observe Regret Learning (Our approach) obtaining superior results with slightly lower volatility and signficantly lower drawdown.}}
  \label{nrlvolatility}
  \vspace{-0.1in}
\end{figure}

Our key contributions can be summarized as follows:
\begin{itemize}
    \item \textbf{Greedy Regret-based Portfolio Strategy}: We introduce a novel portfolio optimization framework based on no-regret learning. The strategy dynamically rebalances assets to minimize hindsight regret, explicitly optimizing risk-adjusted performance. This approach yields an adaptive strategy with superior practical performance.
    \item \textbf{LLM-Guided Clustering and Hedging:} We integrate Large Language Model insights into the portfolio construction pipeline. In particular, we prompt an LLM with market sentiment data (such as CNN’s Fear \& Greed Index) to identify promising sectors and recommend complementary hedges. This sentiment-driven clustering of assets provides a systematic link between behavioral finance signals and allocation decisions. Empirically, the LLM-guided hedging improved risk-adjusted returns substantially – boosting Sharpe ratios by up to 63\% in certain configurations – demonstrating that LLMs can augment human intuition in low-frequency asset management.
    \item \textbf{Empirical Outperformance:} We evaluate the framework on two distinct asset universes: a simple “COCKROACH” mix of cash, gold, bonds, and equities, and a complex set of 51 sector-based ETFs. Across both, our “regret-free” portfolios consistently outperform traditional benchmarks like the S\&P 500 (SPY) and gold. For instance, in the diversified ETF universe, our strategy achieved +68.6\% higher cumulative returns than SPY with 47\% lower maximum drawdown, as shown in Figure \ref{nrlperformance} and \ref{nrlvolatility}. Notably, these gains are attained with a modest quarterly rebalancing frequency – common in index funds – indicating the approach is feasible and cost-effective. Extensive sensitivity analyses (varying rebalancing periods, look-back windows, and sentiment filters) further confirm the strategy’s robust performance across market conditions, with stable or improved returns and controlled risk in each scenario.
\end{itemize}

Overall, our framework of no-regret updates with lightweight sentiment conditioning can produce portfolios that consistently outperform standard benchmarks. This work opens new avenues for integrating online learning with behavioral signals in practical, low-frequency asset management.

\section{Related Work}
\label{sec:related_work}
Portfolio optimization has its foundations in Markowitz's mean variance framework. \cite{Markowitz1952}, which formalized the risk–return trade-off. To better capture downside risks and model uncertainty, alternative measures such as Mean Absolute Deviation (MAD) \cite{Konno1991} and Conditional Value-at-Risk (CVaR) \cite{Rockafellar2000} emerged, alongside robust optimization methods to address computational infeasibility \cite{Goldfarb2003} and noisy data\cite{Mulvey1995}. Sparse or cardinality-constrained models further addressed real-world constraints such as transaction costs and portfolio simplicity \cite{Brodie2009,DeMiguel2009}. However, these static methods struggle to adapt to shifting market dynamics.

Online portfolio optimization frameworks, particularly no-regret algorithms, overcome this limitation by dynamically adjusting to data streams while ensuring performance close to the best fixed strategy in hindsight. \cite{Cover1991,Hazan2007}. 
Pioneering examples include the Universal Portfolio \cite{Cover1991} and Hedge algorithm \cite{Helmbold1998}, 
which asymptotically match optimal constant rebalanced portfolios. Yet, most focus on cumulative return rather than risk-adjusted metrics like Sharpe or Calmar ratios, which are crucial for risk-sensitive investors \cite{EvenDar2006}. Our method builds on this literature by explicitly optimizing such risk-aware objectives within a no-regret learning framework.

Parallel advances in deep and reinforcement learning (RL) have enabled data-driven asset allocation. Deep learning models capture nonlinear return dynamics \cite{Fischer2018,Gu2020}, 
while RL methods learn adaptive allocation strategies through reward-based feedback. \cite{Moody1998,Jiang2017}. 
However, these models often lack interpretability and overfit to noise. In contrast, our approach fuses the adaptivity of these methods with the robustness and transparency of regret minimization.

Behavioral finance further motivates our work by highlighting investor biases such as regret aversion and the longshot bias \cite{Bell1982,Thaler1988}. These preferences drive demand for strategies with built-in risk control and downside protection \cite{Shefrin2000}. Sentiment indicators, especially those capturing fear and greed, offer behavioral context that can inform such strategies.

Recent developments in NLP and Large Language Models (LLMs) make it possible to extract high-level sentiment and thematic signals from text data \cite{Yang2020,Wu2023}. 
Sentiment indices like CNN’s Fear \& Greed Index are increasingly used to anticipate regime shifts, \cite{Baker2007, FarrellOConnor2025, Gomez-Martinez2023}, yet most applications are heuristic or loosely coupled with optimization engines. Our contribution integrates LLM-driven clustering and hedging signals directly into the portfolio generation pipeline, linking behavioral sentiment to allocation in a structured and systematic way.

By combining classical optimization, no-regret learning, deep learning insights, behavioral priors, and LLM-guided sentiment processing, our framework produces a portfolio strategy that is both adaptive and robust, designed to minimize regret in both a theoretical and psychological sense.
\section{Methodology}
This section outlines the dataset design, preprocessing pipeline, and evaluation setup used to assess our proposed no-regret portfolio optimization framework. We describe how asset universes of varying complexity are constructed, how backtest integrity is maintained, and which performance metrics are used to evaluate returns and risks across different market regimes.

\subsection{Asset Universe Construction and Data Sources}
To evaluate our proposed no-regret learning approach for portfolio allocation, we construct two distinct asset universes of increasing complexity to demonstrate the generalizability of our method across diverse investment settings.

\textbf{(i) COCKROACH:} A minimalist asset universe comprising Cash, Gold, 10-Year U.S. Treasury Bonds, and the SPY ETF. This configuration serves as a foundational testbed to illustrate the stability and baseline performance of our method. Daily closing prices were obtained from \url{investing.com} (Cash and Bonds), World Gold Council (Gold), and Yahoo Finance (SPY). % briefly justify why this website

\textbf{(ii) SECTOR-ETFs:} A diversified universe including 51 exchange-traded funds (ETFs) representing 20 economic sectors, supplemented by Cash, Gold, 10-Year Bonds, and SPY. ETFs were sourced from Vanguard, iShares, and SPDR families, with historical data obtained from Yahoo Finance. This design allows for sector-level expressiveness in allocation.

In addition, we incorporate the \textit{Fear and Greed Index}\footnote{Available at \url{https://github.com/gman4774/Fear_and_Greed_Index}} to support dynamic clustering selection within the LLM-guided pipeline.

\subsection{Backtesting Period}
To ensure a meaningful evaluation, we simulate each portfolio strategy over realistic historical periods that reflect diverse macroeconomic conditions. Our chosen timeframes span multiple boom-bust cycles, inflationary shocks, and monetary policy regimes.

\begin{itemize}
    \item \textbf{COCKROACH:} January 3, 2000 – May 31, 2024
    \item \textbf{SECTOR-ETFs:} January 1, 2011 – May 31, 2024
\end{itemize}

The shorter timeframe for SECTOR-ETFs reflects the later introduction of many constituent ETFs, as well as the availability of the Fear and Greed Index, which we used extensively.

A key challenge is the non-uniform listing durations. We address this through the following preprocessing pipeline:
\begin{enumerate}
    \item Remove universal non-trading days (e.g., weekends, market holidays).
    \item Mark asset-specific entry and exit dates.
    \item Represent pre-entry and post-exit days with a placeholder value of $-1$.
    \item Forward-fill missing values due to intermittent non-trading days.
\end{enumerate}

At allocation time, assets with a price of $-1$ on any given day are excluded from optimization. If an asset is delisted while held in the portfolio, it is liquidated to cash, and subsequently blacklisted from future allocations. This ensures consistent matrix dimensions while preserving realism in trading constraints, streamlining further processing on our dataset.

\subsection{Evaluation Benchmarks and Metrics}

To assess the effectiveness of our portfolio strategy, we compare it against traditional buy-and-hold benchmarks and quantify both raw and risk-adjusted returns. This provides a standardized framework for evaluating robustness, capital preservation, and upside capture, and allows us to compare our proposed framework to standard market benchmarks.

We benchmark performance against three passive strategies:
\begin{itemize}
    \item \textbf{SPY Buy-and-Hold}
    \item \textbf{10-Year Treasury Bonds Buy-and-Hold}
    \item \textbf{Gold Buy-and-Hold}
\end{itemize}

Note that these benchmarks are invariant across asset universes and only depend on the backtest timeframe.

The following six metrics are used to evaluate all strategies:
\begin{itemize}
    \item \textbf{Annualized Return} — Average yearly percentage gain over the evaluation period.
    \[
    R_{\text{annual}} = \left( \prod_{t=1}^T (1 + r_t) \right)^{\frac{1}{T_{\text{years}}}} - 1
    \]
    
    \item \textbf{Volatility} — Standard deviation of returns, indicating total risk.
    \[
    \sigma = \sqrt{\frac{1}{T - 1} \sum_{t=1}^T (r_t - \bar{r})^2}
    \]
    
    \item \textbf{Sharpe Ratio} — Risk-adjusted return per unit of total volatility (using 10Y bond returns \(r_f\) as the risk-free rate).
    \[
    \text{Sharpe} = \frac{\bar{r} - r_f}{\sigma}
    \]
    
    \item \textbf{Sortino Ratio} — Risk-adjusted return per unit of downside volatility.
    \[
    \text{Sortino} = \frac{\bar{r} - r_f}{\sigma_d}, \quad \sigma_d = \text{std}\left(r_t \mid r_t < r_f\right)
    \]
    
    \item \textbf{Maximum Drawdown} — Largest observed peak-to-trough decline in portfolio value.
    \[
    \text{Max Drawdown} = \max_{t \in [1,T]} \left( \frac{\max_{\tau \in [1,t]} V(\tau) - V(t)}{\max_{\tau \in [1,t]} V(\tau)} \right)
    \]
    
    \item \textbf{Calmar Ratio} — Annualized return divided by maximum drawdown.
    \[
    \text{Calmar} = \frac{R_{\text{annual}}}{\text{Max Drawdown}}
    \]
\end{itemize}

These metrics jointly capture both return and risk-adjusted performance, enabling a robust comparison of traditional and no-regret allocation methods across asset universes

We also compare the performance of our framework against other established approaches like Mean Absolute Deviation \cite{Konno1991} and Hedge \cite{Helmbold1998}.

\subsection{Architecture}
\label{sec:architecture}

\begin{figure*}[t]
  \centering
  \includegraphics[width=\textwidth]{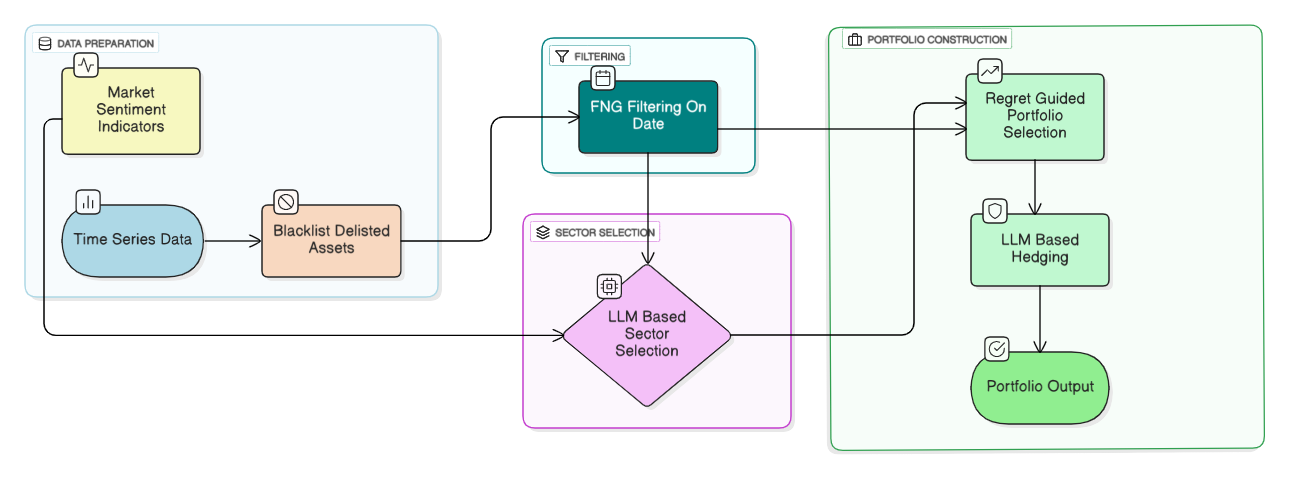}
  \caption{\small Framework modules: (1) Data Prep, (2) Sentiment gating, (3) LLM hedging, (4) Regret-driven allocation.}
  \label{frameworkdiagram}
  \vspace{-0.1in}
\end{figure*}

Our proposed framework models portfolio allocation as a repeated decision-making problem, where an agent dynamically selects portfolio weights over time in response to evolving market conditions. 
As seen in Figure \ref{frameworkdiagram}, the framework has four components:

\begin{enumerate}
    \item \textbf{Data Preparation:} blacklist delisted tickers and align sentiment indices with prices.
    \item \textbf{Sentiment Filter:} a binary gate using CNN’s Fear \& Greed Index, triggering either rebalancing or full liquidation during extreme sentiment.
    \item \textbf{LLM Module:} interprets sentiment to propose sector hedges (not primary allocation). Clustering was tested but found weaker; hedging yielded consistent drawdown reduction.
    \item \textbf{Regret-Guided Allocation:} FTL-inspired rebalancing with candidate action perturbations.
\end{enumerate}

\paragraph{Sentiment filtering Module.}
This component serves as a rebalancing filter, driven by market sentiment. It receives as input a market sentiment index (e.g., CNN’s Fear and Greed Index) and outputs a binary decision: either to proceed with rebalancing the portfolio or to liquidate it entirely. This mechanism acts as a coarse-grained risk control measure, enforcing conservative action during extreme sentiment conditions. Further implementation details are provided in Section~\ref{sec:params-design-choices}.

\paragraph{LLM-Based Sector Selection and Hedging Module.}
This module leverages a language model to interpret recent market sentiment trends and propose sector-level adjustments. Specifically, it suggests promising sectors for inclusion and complementary sectors for hedging, aiming to improve diversification and stability. Rather than integrating directly into the core regret-minimization loop, it serves as an upstream filter that constrains the action space by dynamically shaping the investable universe and risk posture. The prompting methodology, voting logic, and filtering heuristics used to operationalize this module are detailed under our dynamic clustering and pseudo-hedging mechanisms in Section~\ref{sec:params-design-choices}.

\paragraph{Regret-Based Portfolio Construction Mechanism.}
At the heart of our framework lies a regret-minimization strategy that governs portfolio construction. This component integrates three key components: a game-theoretic formulation, a no-regret learning objective, and a systematic action-generation procedure that explores perturbations around a baseline allocation.

\subsubsection{Game-Theoretic Formulation}

We model the portfolio optimization task as a single-agent game where the agent interacts with a dynamic financial environment by selecting portfolio weights over time. At each timestep \( t \), the agent chooses an allocation \( \mathbf{a}(t) = [a_1(t), a_2(t), \dots, a_n(t)] \) over \( n \) assets. Each weight \( a_i(t) \in [0, 1] \), and the allocation satisfies the constraint:

\[
\sum_{i=1}^{n} a_i(t) = 1.
\]

The objective is to optimize the portfolio based on a selected performance criterion. We define a set of admissible metrics:

\[
\mathcal{O} = \{\text{Annualized Returns}, \text{Sharpe Ratio}, \text{Calmar Ratio}, \text{Sortino Ratio}\}.
\]

At training time, a single objective \( O \in \mathcal{O} \) is chosen, and the agent's goal becomes:

\begin{equation}
\max_{\mathbf{a}(t)} O(\mathbf{a}(t)) \quad \text{or} \quad \min_{\mathbf{a}(t)} O(\mathbf{a}(t)),
\end{equation}

depending on the metric’s orientation (e.g., minimizing volatility vs. maximizing returns).

\subsubsection{Regret Guided Learning Objective}

To encourage long-term optimality, we adopt a \textit{regret guided learning} paradigm. The goal is to ensure that the agent’s cumulative performance approaches that of the best fixed decision in hindsight. In other words, the algorithm should incur minimal regret for not having selected a single optimal allocation from the beginning. However, as we will see, we deviate from the guarantees provided by the original objective.

Let \( \mathcal{A} \) denote the set of all valid allocations (i.e., portfolio vectors) such that each asset weight \( a_i \in [0,1] \) and the constraint \( \sum_{i=1}^{n} a_i = 1 \) holds. Let \( \mathbf{a}^* \in \mathcal{A} \) be the best fixed allocation in hindsight. Define a metric \( M(\mathbf{a}, t) \) that evaluates the performance (e.g., return, Sharpe ratio) of an allocation \( \mathbf{a} \) at time \( t \). The regret over a time horizon \( T \) is defined as:

\begin{equation}
\text{Regret}(\mathbf{a}^*, T) = \frac{1}{T} \sum_{t=1}^{T} \left[ M(\mathbf{a}^*, t) - M(\mathbf{a}(t), t) \right],
\end{equation}

where \( \mathbf{a}(t) \) is the allocation selected by the agent at time \( t \). A strategy is said to be \textit{no-regret} if the average regret vanishes asymptotically:

\begin{equation}
\lim_{T \to \infty} \text{Regret}(\mathbf{a}^*, T) = 0.
\end{equation}

% \begin{figure}
%   \centering
%   \includegraphics[width=0.35\textwidth]{fig/block_diagram.drawio.png}
%   \caption{\small{The framework comprises three modules: (1) Regret-Based Portfolio Selection via no-regret learning; (2) F\&G-Based Evaluation for sentiment-driven rebalancing; and (3) LLM-Based Sector Selection and Hedging for downside protection.}}
%   \label{frameworkdiagram}
%   \vspace{-0.1in}
% \end{figure}

This formulation ensures that the agent's cumulative performance matches that of the optimal fixed portfolio as the time horizon grows. No-regret learning serves as a foundational concept in online learning theory, offering strong performance guarantees even in adversarial or non-stationary environments. %\cite{cesa2006prediction, hazan2016introduction}. 
In the context of portfolio selection, no-regret algorithms aim to match the performance of the best constant rebalanced portfolio in hindsight. %\cite{cover1991universal, agarwal2006algorithms}. 

Classical approaches like the Hedge algorithm %\cite{freund1997decision}
achieve sublinear regret by maintaining and updating a distribution over expert strategies or assets. Our approach is conceptually similar, but deviates in the policy update mechanism: rather than maintaining weighted mixtures, we select the policy that has achieved the highest absolute performance so far. This can be viewed as a variant of the \textit{“follow-the-leader”} strategy, where the algorithm greedily chooses the best-performing allocation in hindsight at each round. However, this greedy approach departs from the principled weighting schemes of classical no-regret algorithms and consequently forfeits the theoretical guarantee of sublinear (vanishing) regret.

Formally, at each rebalancing step, the agent selects \( \mathbf{a}(t) \in \mathcal{A} \) such that:

\begin{align}
0 \leq a_i(t) \leq 1 \quad &\forall i, t, \\
\sum_{i=1}^{n} a_i(t) = 1 \quad &\forall t.
\end{align}

This policy selection scheme adheres to the same regret minimization principle, but is more deterministic and performance-driven.

Although our evaluation is performed retrospectively on historical data (i.e., offline), the algorithm itself follows an online learning protocol, making sequential allocation decisions at each timestep using only information available up to that point. At each timestep \( t \), the agent makes allocation decisions based solely on past information, with no access to future data and without relying on parameters pre-trained on the test set, thereby respecting the constraints of online learning and avoiding data leakage.

\subsubsection{Action Set Generation}

We generate candidate actions by perturbing a baseline allocation in a structured manner. The baseline allocation is initially set to an all cash portfolio. Then at each rebalancing step, the baseline allocation is the existing allocation at that timestep before rebalancing. At each timestep, a single asset’s allocation is modified within a predefined interval, and the remaining allocations are adjusted to preserve feasibility. This allows for controlled exploration of the allocation space.

The procedure is outlined below:

\begin{algorithm}[H]
\caption{Generate Action Scenarios}
\begin{algorithmic}[1]
\State Initialize baseline allocation $\mathbf{a}_{\text{base}}$
\For{each asset $i$ in $\mathbf{a}_{\text{base}}$}
    \For{allocation $a_i'$ from $a_{\min}$ to $a_{\max}$}
        \State Compute $\Delta a_i \gets a_i' - a_i$
        \State Adjust remaining allocations to maintain $\sum_j a_j = 1$
        \If{All allocations satisfy $0 \leq a_j' \leq 1$}
            \State Add $\mathbf{a}'$ to action set $\mathcal{A}$
        \EndIf
    \EndFor
\EndFor
\State \Return Action set $\mathcal{A}$
\end{algorithmic}
\end{algorithm}

Key components of this setup include:
\begin{itemize}
    \item \( a_{\min} \) and \( a_{\max} \): lower and upper bounds for allowable asset weight perturbation.
    \item Adjustments to non-target assets ensure the sum-to-one constraint is preserved.
    \item Only one asset’s allocation is altered at each timestep, enabling structured and interpretable deviations from the baseline.
\end{itemize}

This action generation scheme supports efficient search within the valid allocation simplex and ensures compliance with budget and allocation constraints.

\subsection{Parameters and Design Choices}
\label{sec:params-design-choices}

We now outline the key hyperparameters and design choices that govern the behavior and flexibility of our no-regret portfolio allocation framework.

\textbf{Rebalancing Frequency} defines the interval at which the agent is allowed to update the portfolio. In our experiments, we use a slow rebalancing schedule (e.g., quarterly) to limit turnover and minimize transaction costs. This makes our framework suitable for semi-passive asset management strategies, such as those employed by pension or mutual funds, rather than for high-frequency or swing trading.

\textbf{Minimum and Maximum Allocations} are constraints placed on the share of portfolio capital allocated to each asset. These enforce diversification by ensuring that no asset can dominate the portfolio or be entirely excluded, unless specifically delisted.

\textbf{Window  or \textit{k}} specifies the lookback horizon over which the performance of candidate allocations is evaluated. This is separate from rebalancing frequency and governs how much historical data is used in generating regret-minimizing policies.

\textbf{Objective} refers to the performance metric that the algorithm is tasked with optimizing. The user may choose from any of the six supported metrics: Annualized Return, Sharpe Ratio, Sortino Ratio, Calmar Ratio.

\textbf{Cluster Filter} allows the user to restrict the asset universe to a subset of sectors. This is particularly relevant in SECTOR-ETFs and SPDR500 universes. If set to ``dynamic," the cluster is determined using an LLM-driven pipeline guided by historical fear and greed data (explained below). Our sector classification includes 20 groups—an extension over the standard 12 market sectors used by Yahoo Finance.

\textbf{Exponential Weighting} determines how asset returns are aggregated over the lookback window. When enabled, more recent days are weighted more heavily to better reflect recent market trends. When disabled, uniform weighting is applied across the window.

\textbf{Percentile Filter} smooths returns by filtering the asset universe to exclude extreme performers. Assets are ranked by their historical performance, and only those within a configurable percentile range (e.g., 30th–70th) are retained. This is motivated by the intuition that mid-performing assets may offer more stable return profiles than either outperformers or laggards.

\textbf{Fear and Greed (F\&G) Filters} include three mechanisms: upper bound, lower bound, and delta-bounded. These restrict rebalancing to market regimes with acceptable sentiment. The delta-bounded filter accepts parameters $(l, h, t)$, which define a permitted range for the change in sentiment between $t$ days prior and the rebalancing date. If no filter is satisfied on a rebalancing day, the portfolio may optionally be converted entirely to cash to limit risk exposure.

\textbf{Uniform Adjustment Mechanism} governs how capital is redistributed when modifying a single asset's weight during action generation. If enabled, any increase or decrease in allocation to one asset is offset proportionally across all other invested assets. Otherwise, the residual is shifted to or from cash. This helps embed structural diversification into the action space.

\textbf{Dynamic Clustering} leverages historical fear and greed data to guide sector selection during rebalancing. For each rebalancing point, a 12-month fear-and-greed time series is fed into an LLM, which is prompted to suggest the three most promising sectors. To mitigate variance in LLM outputs, we repeat the prompt five times and take the most frequently recommended sectors. This replaces the standard “static” sector filtering. Sector tags are manually assigned to each asset and span 20 categories for finer control.

\textbf{Pseudo-Hedging Mechanism} is applied after dynamic clustering. The LLM is prompted (without any reference to dates) to suggest sector-level hedges for each selected cluster. We avoid equity-level hedging due to potential LLM biases toward frequently mentioned tickers (e.g., AAPL vs. lesser-known tickers like VLTO). This mechanism is designed to reduce volatility and improve Sharpe ratio by adding non-correlated sectors to the portfolio.

\subsection{Backtesting Mechanics}

We developed a custom backtesting framework to simulate the agent’s behavior under realistic constraints and assess performance across different asset universes.

The framework takes as input a CSV file where each column corresponds to an asset’s daily closing price. The date column is resampled using the rebalancing frequency to generate valid rebalancing points. On each of these dates, the agent is allowed to reallocate the portfolio according to the no-regret policy.

\begin{table*}[htbp]
\centering
\resizebox{\textwidth}{!}{%
\begin{tabular}{lllllllllllllll}
\toprule
\textbf{Objective} & \textbf{Rebal.} & \textbf{Sector} & \textbf{F\&G} & \textbf{F\&G} & \textbf{k} & \textbf{Pct.} & \textbf{Lookback} & \textbf{LLM} & \textbf{Return} & \textbf{Vol.} & \textbf{Draw-} & \textbf{Sharpe} & \textbf{Sortino} & \textbf{Calmar} \\
& \textbf{Freq.} & \textbf{Cluster} & \textbf{Thresh.} & \textbf{Change} & & \textbf{Thresh.} & \textbf{Days} & \textbf{Hedge} & \textbf{(\%)} & & \textbf{down} & \textbf{Ratio} & \textbf{Ratio} & \textbf{Ratio} \\
\midrule
Calmar & Q & All & 10--90 & 20\% / 5d & 30 & - & - & N & \textbf{18.91} & 0.15 & 0.16 & \textbf{1.03} & \textbf{1.346} & \textbf{1.1965} \\
Return & Q & All & 10--90 & 20\% / 5d & 30 & - & - & N & 16.81 & 0.17 & 0.22 & 0.80 & 1.0816 & 0.7504 \\
Sortino & Q & All & 10--90 & 20\% / 5d & 30 & - & - & N & 16.04 & 0.15 & \textbf{0.15} & 0.89 & 1.1916 & 1.0502 \\
Sortino & Q & All & 10--90 & 20\% / 5d & 10 & - & - & N & 15.24 & 0.15 & 0.19 & 0.84 & 1.1333 & 0.8035 \\
Calmar & Y & Dynamic (LLM) & 10--90 & - & 90 & 50 & 30 & N & 11.46 & 0.14 & 0.28 & 0.61 & 0.7477 & 0.4063 \\
Return & Y & Dynamic (LLM) & 10--90 & 20\% / 5d & 30 & - & - & N & 11.54 & 0.13 & 0.23 & 0.64 & 0.6317 & 0.4936 \\
Return & Q & Dynamic (LLM) & 10--90 & 20\% / 5d & 120 & - & - & Y & 13.08 & \textbf{0.12} & 0.18 & 0.83 & 1.0335 & 0.739 \\
Return & Q & Dynamic (LLM) & 10--90 & - & 60 & - & - & Y & 12.43 & \textbf{0.12} & 0.22 & 0.77 & 0.9832 & 0.5704 \\
\midrule
\multicolumn{15}{l}{\textbf{Benchmark Reference}} \\
\midrule
\textbf{Online Mean Absolute Deviation\cite{Konno1991}} & -- & -- & -- & -- & -- & -- & -- & -- & 13.56 & 0.14 & 0.24 & 0.74 & 0.95 & 0.58 \\
\textbf{Hedge (Multiplicative Weights)\cite{Helmbold1998}} & -- & -- & -- & -- & -- & -- & -- & -- & 17.09 & 0.19 & 0.26 & 0.76 & 1.00 & 0.65 \\
\textbf{SPY} & -- & -- & -- & -- & -- & -- & -- & -- & 11.21 & 0.17 & 0.34 & 0.47 & 0.58 & 0.33 \\
\textbf{Gold} & -- & -- & -- & -- & -- & -- & -- & -- & 3.91 & 0.16 & 0.45 & 0.06 & 0.08 & 0.09 \\

\bottomrule
\end{tabular}%
}
\caption{Performance of various portfolio strategies under different configurations in a grid search from January 2011 to June 2024.}
\label{tab:strategy-performance}
\end{table*}

To ensure economic realism, we incorporate daily price tracking of each asset, including cash (held at \$1 per dollar) and 10-year Treasury bonds (which are priced from yield curves assuming zero-coupon stripping). Since bonds are not held to maturity, we calculate their market price on each day to enable trading.

Delisted assets are handled by forward-filling their price until official delisting, after which they are added to a blacklist and excluded from future portfolios. On rebalancing dates, if the asset’s price is marked as $-1$, it is discarded from any possible portfolio allocations.

Sentiment-aware components are triggered at each rebalancing step. If any F\&G filters are active, the framework retrieves the current index and optionally the index from $t$ days ago. These are used to determine whether rebalancing is permitted. If dynamic clustering is enabled, the 12-month sentiment time series is passed to the LLM to determine a filtered subset of sectors. If hedging is enabled, the LLM is queried for complementary sectors and these are added to the universe.

The no-regret learning algorithm is then executed. It generates a constrained set of candidate allocations (e.g., by perturbing one asset at a time) and selects the one with the best historical performance under the chosen objective. Allocations obey constraints such as minimum/maximum exposure per asset, and changes are conservatively scoped to reduce transaction risk.

Each allocation results in a simulated liquidation of the existing portfolio and reallocation into the new one. Daily portfolio value is then tracked until the next rebalancing date.

We also incorporate transaction costs using a flexible fee model that supports both per-share fees and percentage-based costs. Fees are subtracted on each rebalancing day to simulate realistic trading conditions.
\section{Results}
\label{sec:results}
The objective of our approach was to develop a quantitative strategy that matches the returns of benchmark funds (such as SPY) while significantly reducing risk and drawdowns. Achieving this balance would make the portfolio attractive to risk-averse investors and fund managers seeking low correlation with the broader market. In essence, a high Sharpe ratio strategy can be leveraged by fund managers to obtain more return per unit of risk, offering a compelling proposition in risk-adjusted performance.

We now analyze performance across multiple asset universes, investigate the impact of core design choices, and assess the role of LLM-guided augmentations.

\subsection{Performance Across Asset Universes}

\begin{figure}[H]
  \centering
  \includegraphics[width=0.45\textwidth]{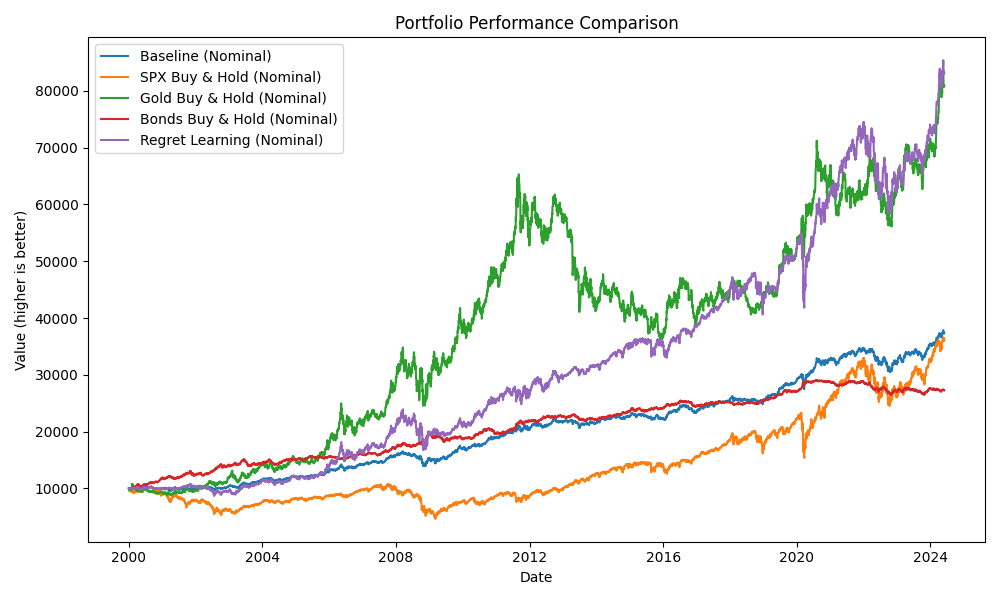}
  \caption{\small{The No Regret Learning strategy (purple) outperforms traditional buy-and-hold benchmarks (SPX, gold, bonds, static baseline) on COCKROACH through adaptive allocation and lower volatility, demonstrating superior long-term performance and robustness.}}
  \label{simpleportfolioreturns}
\end{figure}

\textbf{COCKROACH Universe:} In the COCKROACH asset universe (comprised of just four diversified assets), NRL achieved performance marginally exceeding both the SPY and Gold in terms of total returns, while maintaining a much steadier growth trajectory, as seen in Figure \ref{simpleportfolioreturns}. This strong performance with only four assets suggests that the model leverages the broad diversification inherent in SPY’s holdings—effectively capitalizing on the fact that SPY represents a basket of strong underlying businesses with solid fundamentals. By relying on SPY (an index fund with internal active management of constituents) along with a few other uncorrelated assets (e.g., Gold), the NRL strategy was able to deliver stock-like returns with significantly lower volatility.

\begin{figure}
  \centering
  \includegraphics[width=0.45\textwidth]{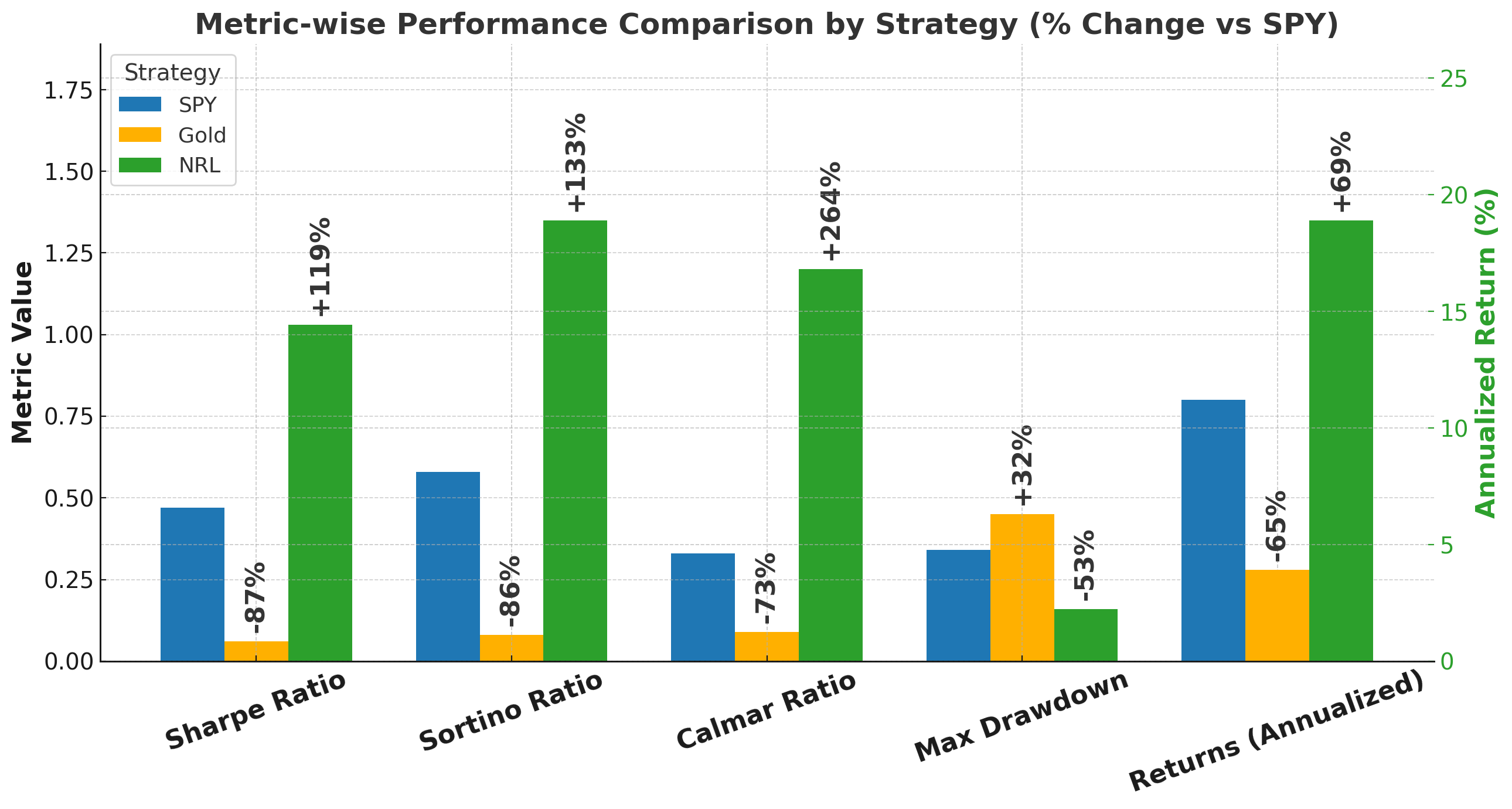}
  \caption{\small{NRL's performance across multiple metrics as compared to SPY (benchmark) from January 2011 to June 2024, using the top configuration shown in Table \ref{tab:strategy-performance}}. (separate scale used for returns)}
  \label{metricplot}
  \vspace{-0.1in}
\end{figure}

\textbf{SECTOR-ETFs Universe:} Building on the above, we expanded the asset universe by allowing NRL to invest in sector-specific ETFs (managed funds representing each market sector), while applying F\&G index-based filter to manage extreme market conditions. On this broader SECTOR-ETFs universe, NRL demonstrated a similar trend of outperformance, as seen in Figure \ref{metricplot}. The NRL-driven portfolio outpaced the SPY benchmark over the same 25-year test period by +68.6\% in cumulative returns and achieved 119\% higher Sharpe ratio—dramatically better risk-adjusted results. Additionally, its Calmar ratio was roughly +262\% higher. Notably, the maximum drawdown of the NRL portfolio in the SECTOR-ETFs test was around 16\% over approximately 25 years, compared to the 34\% maximum drawdown experienced by SPY in the same timeframe. This reduction in drawdown highlights NRL’s ability to mitigate downside risk while still capturing upside gains.

\subsection{Key Factors Influencing Strategy Performance}
Below we summarize the key influences that had an impact over the performance of NRL strategy:

\textbf{Rebalancing Frequency:} The frequency of portfolio rebalancing emerged as an important factor. Among all the parameter configurations that outperformed SPY, 71\% employed a quarterly rebalancing schedule. This suggests that medium-term market trends often persist on the order of a few months. Rebalancing approximately every three months may capture these trends effectively, whereas rebalancing less frequently risks the portfolio drifting away from evolving opportunities. Quarterly rebalancing thus appears to strike a desirable balance between being responsive to market changes and avoiding over-trading.

\textbf{Optimization Objective:} We experimented with different objective functions for the NRL algorithm (e.g., optimizing for raw return vs. optimizing for risk-adjusted measures). The majority of SPY-beating configurations (about 50\%) were achieved by optimizing for pure return, while roughly 25\% optimized for the Calmar ratio and 25\% for the Sharpe ratio. Interestingly, although maximizing returns yielded many profitable configurations, the highest quality performance metrics were attained when optimizing for Sharpe or Calmar ratios. This suggests that while it is easier to find a profitable configuration using return as an objective, optimizing for risk/reward tradeoffs often delivers superior long-term performance and robustness.

\textbf{Asset Universe Breadth:} The number of assets and the use of performance-based filtering also affected results. Although we initially explored percentile filtering to include only “top” performing assets and reduce noise, our results showed that not applying percentile filters generally led to higher Sharpe ratios. We interpret this as evidence that a broader asset universe allows the model more flexibility in rebalancing and capital allocation. With more available assets, the agent can better adapt to emerging opportunities and construct more diversified portfolios, ultimately improving performance stability.

% (graphs and plots will accompany this section)

\subsection{LLM-Based Augmentations and Signal Conditioning}

Beyond standard parameter tuning, we explored integrating large language model (LLM)-driven mechanisms into portfolio construction, focusing on dynamic clustering, hedging, and F\&G-index-based rebalancing strategies to embed macro-level sentiment into asset allocation.

Experiments using LLM-driven dynamic sector selection—guided by temporal F\&G index patterns—did not outperform the full asset universe. Dynamic clustering configurations exhibited cumulative returns and Sharpe ratios averaging 34\% and 37\% lower, respectively, than the baseline (configuration Z). This performance decline appears due to LLM limitations in reliably identifying predictive sector clusters solely from fear-and-greed data, particularly during volatile market regimes.

Conversely, LLM-based hedging significantly improved risk-adjusted returns. Combined with dynamic clustering, LLM-generated hedges increased Sharpe ratios by 31\% on average, with quarterly rebalancing achieving even greater improvements (63\%) relative to non-hedged configurations. These findings indicate that while LLMs struggle with high-return sector identification, they effectively pinpoint complementary or risk-offsetting sectors, enhancing robustness.

Lookback window length for signal evaluation emerged as critical. With dynamic clustering enabled, longer windows consistently boosted Sharpe ratios, notably with a 120-day window outperforming a 10-day window by 17\%. Without clustering, Sortino ratios increased nearly linearly beyond 60 days, while shorter windows displayed erratic performance and a local optimum near 30 days under quarterly rebalancing.

We also examined F\&G index movements as rebalancing signals. Applying a volatility filter—suppressing rebalancing if the F\&G index changed more than a threshold over recent trading sessions consistently improved performance. A threshold of a 20-point absolute change over five days optimally filtered short-term sentiment noise while preserving timely tactical adjustments.

Finally, imposing absolute-level constraints on the F\&G index provided additional risk control. Blocking trades when the index fell below 10 or exceeded 90 optimally balanced market participation with drawdown protection. Tighter constraints (e.g., limiting trades to a 30–70 index range) substantially lowered returns and Sharpe ratios, likely due to missed high-momentum opportunities. These results indicate that moderate exposure to extreme sentiment regimes can be advantageous if accompanied by effective risk management.

\section{Limitations and Future Work}
\label{sec:limitations}
The primary limitation of our approach lies in the unpredictability of performance relative to the chosen optimization objective, whether maximizing return, Sharpe ratio, Calmar ratio, or another metric. For instance, configurations optimizing for Calmar ratio frequently yielded the highest returns, which is counterintuitive given that Calmar emphasizes drawdown mitigation rather than absolute performance. This misalignment introduces uncertainty for end-users, especially risk-averse investors requiring a reliable relationship between objectives and outcomes.

Another constraint is reliance on fixed-frequency rebalancing schedules. While periodic rebalancing is effective under stable conditions, it may prove suboptimal during heightened volatility or structural market shifts. Without adaptive mechanisms, this rigidity can result in missed opportunities or amplified losses, as during extended downturns like the 2022 market crash. Future work could explore event-driven rebalancing triggered by significant market signals or volatility thresholds, improving responsiveness to adverse conditions.

The role of LLMs was intentionally limited, primarily focusing on hedging. Although LLMs showed utility in identifying risk-offsetting exposures, their potential for more granular, asset-level or sector-level decisions remains underexplored. Given the infrequent rebalancing constraints, micro-level applications of LLMs were minimally tested. Future research could investigate fine-grained uses, such as interpreting real-time news sentiment to dynamically adjust sector weights or allocations in response to unfolding events.
\section{Conclusion}
\label{sec:conclusion}
In this work, we introduced a novel portfolio optimization strategy leveraging no-regret learning combined with Large Language Model-driven smart clustering and hedging. Our empirical results demonstrate that this integrated framework significantly outperforms traditional buy-and-hold benchmarks across various market conditions and asset universes, delivering superior risk-adjusted returns with reduced drawdowns. While our method shows robustness and adaptability, limitations such as fixed-frequency rebalancing and LLM predictive constraints highlight opportunities for future research. Overall, our approach presents a compelling framework for investors seeking reliable, risk-aware, and adaptive asset management.

\bibliographystyle{ACM-Reference-Format}
\bibliography{ref}

\end{document}